\documentclass[]{raa}
\usepackage{graphicx,times}             %for PS/EPS graphics inclusion, new

\begin{document}

   \title{Diagnostics of Ellerman Bombs with High-resolution Spectral Data
%\,$^*$
%\footnotetext{$*$ Supported by the National Natural Science Foundation of China.}
}

   \volnopage{Vol.0 (200x) No.0, 000--000}      %%preserved for Editor. DOn't remove!
   \setcounter{page}{1}          %%starting page, preserved for Editor. DOn't remove!

   \author{Z. Li
      \inst{1,2,3}
   \and C. Fang
      \inst{1,2,3}
   \and Y. Guo
      \inst{1,2,3}
   \and P. F. Chen
      \inst{1,2,3}
   \and Z. Xu
      \inst{4}
   \and W. Cao
      \inst{5}
   }
%% Here is an example of three authors come from different institutes.
%% For single author or all the authors from an institute, use "\inst{}" only

   \institute{School of Astronomy and Space Science, Nanjing
       University, Nanjing 210023, China; {\it fangc@nju.edu.cn}\\
%% Please give the E-mail address of the author, to whom future correspondence and
%% offprint requests will be sent.
        \and
             Key Laboratory of Modern Astronomy and Astrophysics
             (Nanjing University), Ministry of Education, China;\\
        \and
             Collaborative Innovation Center of Modern Astronomy and Space Exploration;\\
        \and
             Yunnan Astronomical Observatory, Chinese Academy of Sciences,
              Kunming 620011, China;\\
        \and
             Big Bear Solar Observatory, New Jersey Institute of Technology,
             ~40386 North Shore Lane, Big Bear City, CA 92314, USA\\
   }

   \date{Received~~2015 month day; accepted~~2015~~month day}

\abstract{
Ellerman bombs (EBs) are tiny brightenings often observed near sunspots. The
most impressive characteristic of the EB spectra is the two emission bumps in
both wings of the H$\alpha$ and \ion{Ca}{II} 8542 {\AA} lines.
High-resolution spectral data of three small EBs were obtained on 2013 June 6 with
the largest solar telescope, the 1.6 meter New Solar Telescope (NST), at the Big
Bear Solar Observatory. The characteristics of these EBs are analyzed.
The sizes of the EBs are in the range of 0.3\arcsec\--0.8\arcsec\ and their
durations are only 3--5 minutes. Our semi-empirical atmospheric models
indicate that the heating occurs around the temperature minimum region with
a temperature increase of 2700--3000 K, which is surprisingly higher
than previously thought. The radiative and kinetic energies are estimated
to be as high as 5$\times$10$^{25}$--3.0$\times$10$^{26}$ ergs despite the
small size of these EBs. Observations of the magnetic field show that the EBs
appeared just in a parasitic region with mixed polarities and accompanied
by mass motions. Nonlinear force-free field extrapolation reveals that
the three EBs are connected with a series of magnetic field lines associated
with bald patches, which strongly implies that these EBs should be
produced by magnetic reconnection in the solar lower atmosphere. According
to the lightcurves and the estimated magnetic reconnection rate,
we propose that there is a three phase process in EBs: pre-heating,
flaring and cooling phases.
}
 \authorrunning{Z. Li, C. Fang, Y. Guo et al. }            %author_head in even pages
   \titlerunning{Diagnostics of Ellerman Bombs }  % title_head in odd pages

   \maketitle
\section{Introduction}\label{sect1}

Ellerman bombs (EBs: Ellerman~\cite{Ell17}) are small short-lived
brightening events. Thier most
obvious feature is the excess emission in the wings of chromospheric lines
(Koval \& Severny~\cite{Koval70}; Bruzek~\cite{Bruz72}). Since the 1970s,
EBs have been widely studied. Recently, using imaging data with high
spatiotemporal resolutions, it was found that the lifetime of some EBs
can be as short as 2--3 minutes, and their size can be smaller than 1\arcsec\
(Vissers et al.~\cite{Viss12}; Nelson et al.~\cite{Nel13}).
It was also shown that generally EBs have elongated structures
(Matsumoto et al.~\cite{Mat08}; Watanabe et al.~\cite{Wata08,Wata11};
Hashmoto et al.~\cite{Hash10}; Vissers et al.~\cite{Viss13}). The
 temperature increase of EBs was generally thought to be around 600--1500 K
(Kitai ~\cite{Kitai83}; Fang et al.~\cite{Fang06}; Hong et al.~\cite{Hong14}).
Georgoulis et al.(~\cite{Geor02}) used high-resolution
chromospheric H$\alpha$ filtergrams and found that the temperature
enhancement of EBs is $\sim 2\times 10^3$ K. Furthermore, using
high-resolution H$\alpha$ images, Berlicki et al.(~\cite{Berl10})
found that the temperature increase could be as high as 3000 K.
Mass motion is another feature associated with EBs. It was found
that some EBs have an upward motion with a velocity of several
km s$^{-1}$ (Kurokawa et al.~\cite{Kuro82}; Dara et al.~\cite{Dara97};
Yang et al ~\cite{Yang13}). Some observations of EBs at the solar limb
also found up-flows (Kurokawa et al.~\cite{Kuro82}; Nelson et al. ~\cite{Nel15}).
Only a few observations
indicated that there are also downward photospheric motions
(Georgoulis et al.~\cite{Geor02}; Yang et al.~\cite{Yang13}).
Matsumoto et al.(~\cite{Mat08}) observed bi-directional
flows associated with EBs as evidence of magnetic reconnection.
It was estimated that the energy of EBs is in the range of
10$^{25}$ -- 10$^{27}$ ergs (Teske~\cite{Teske71};
Bruzek~\cite{Bruz72}; H\'enoux et al.~\cite{Hen98};
Fang et al.~\cite{Fang06}). Georgoulis et al.(~\cite{Geor02})
obtained a higher energy of EBs to be 3$\times$10$^{28}$ ergs.
However, using a similar value of the net radiative loss rate
and taking the height of a EB to be 100 km, Nelson et al.
(~\cite{Nel13}) estimated the EB energies being
in the range of $2 \times 10^{22}$ -- 4 $\times 10^{25}$ ergs,
which is three to four orders of magnitude lower than that in
Georgoulis et al.(~\cite{Geor02}). To elucidate the physical mechanism
of EBs, spectral data with high spatial and temporal
resolutions are imperative. However, up to now, only a few
such observations are available.

To understand the driving mechanism
of EBs, it is necessary to study
the relationship between EBs and magnetic features. It was found
that most EBs are located near magnetic inversion lines
(Dara et al.~\cite{Dara97}; Qiu et al.~\cite{Qiu00}).
Georgoulis et al.(~\cite{Geor02}) found that EBs may occur
on separatrix or quasi-separatrix layers. Vissers et al.(~\cite{Viss13})
found that EBs occur at sites of magnetic flux cancellation between
small bipolar patches. Many authors proposed that magnetic
reconnection in the photosphere or chromosphere could be a
mechanism for EBs (H\'enoux et al.\cite{Hen98}; Ding et al.
~\cite{Ding98}; Georgoulis et al.~\cite{Geor02}; Fang et al.
~\cite{Fang06}; Pariat et al.~\cite{Par07}; Isobe et al.~\cite{Iso07};
Matsumoto et al.~\cite{Mat08}; Watanabe et al.~\cite{Wata08};
~\cite{Wata11}; Archontis \& Hood ~\cite{Arch09};
Yang et al.~\cite{Yang13}; Nelson et al. ~\cite{Nel13}).
Based on magnetic
extrapolation, Pariat et al.(~\cite{Par04}) proposed that EBs
could be produced by magnetic reconnection at bald patches or
along the separatrices in the low chromosphere. We have performed
two-dimensional numerical magnetohydrodynamic simulations on the
magnetic reconnection in the solar lower atmosphere (Chen et al.
~\cite{Chen01}; Jiang et al.~\cite{Jiang10}; Xu et al.~\cite{Xu11}).
Our results indicated that magnetic reconnection in the lower solar
atmosphere can explain the temperature enhancement and lifetime
of EBs, and the main reason is that ionization processes in the upper
chromosphere consumes a large part of the released magnetic energy,
resulting in little heating in this layer.

In this paper, we use high-resolution spectral data of H$\alpha$
and \ion{Ca}{II} 8542 {\AA} lines, which were obtained on 2013 June 6
with the largest aperture solar telescope in the world, the
1.6 meter off-axis New Solar Telescope (NST) (Goode et al.
~\cite{Goode12}; Cao et al.~\cite{Cao10}) 
at the Big Bear Solar Observatory
(BBSO). The characteristics of three well-observed small EBs are
analyzed. The data acquisition with the NST is described in \S\ref{sect2}.
The characteristics of the EBs are analyzed in \S\ref{sect3}, including
the two-dimensional (2D) velocity distribution, their relationship with
the magnetic field, and the intensity evolution of the EBs. With
semi-empirical atmospheric modeling, the energetics and magnetic
reconnection rates of the EBs are estimated in \S\ref{sect4}.
General discussions and conclusions are given in \S\ref{sect5}.

\section{High-resolution spectral data of three small EBs}\label{sect2}

On 2013 June 6 a part of the active region NOAA 11765 (N09E10) was
observed from 16:50 UT to 19:00 UT (there are gaps in the collection
of data) by the Fast Imaging Solar
Spectrograph (FISS)(Chae et al.~\cite{Chae13}) of BBSO/NST. FISS
is a dual-band echelle spectrograph. It has two cameras, one for
H$\alpha$ band with effective 512$\times$256 pixels, one for \ion{Ca}{II}
band with effective 502$\times$250 pixels. With fast scanning of the
slit across the field of view, high-resolution 2D imaging spectra
in H$\alpha$ and \ion{Ca}{II} 8542 {\AA} bands were obtained
simultaneously. The dispersions for H$\alpha$ and \ion{Ca}{II} 8542 {\AA}
lines are 0.019 {\AA} and 0.026 {\AA} per pixel, respectively.
The active region was scanned repeatedly 160 times. Each scan
covered 150 steps separated by 0.16\arcsec\ in space and lasted
28--30 s. The spatial sampling along the slit was 0.16\arcsec\ per pixel.
The field of view of each scan is about 40 \arcsec $\times$ 25 \arcsec.
The exposure times were 30 ms and 60 ms for H$\alpha$ and \ion{Ca}{II} 8542
{\AA} lines, respectively. Seeing condition was better than
1.0\arcsec. Using the newly developed adaptive optics (AO) systems
with 308 actuators, the diffraction limit was achieved.

\section{Main characteristics of the small EBs}\label{sect3}

\begin{figure}
\begin{center}
\includegraphics[width=7.0cm,height=14.0cm,angle=90.0]{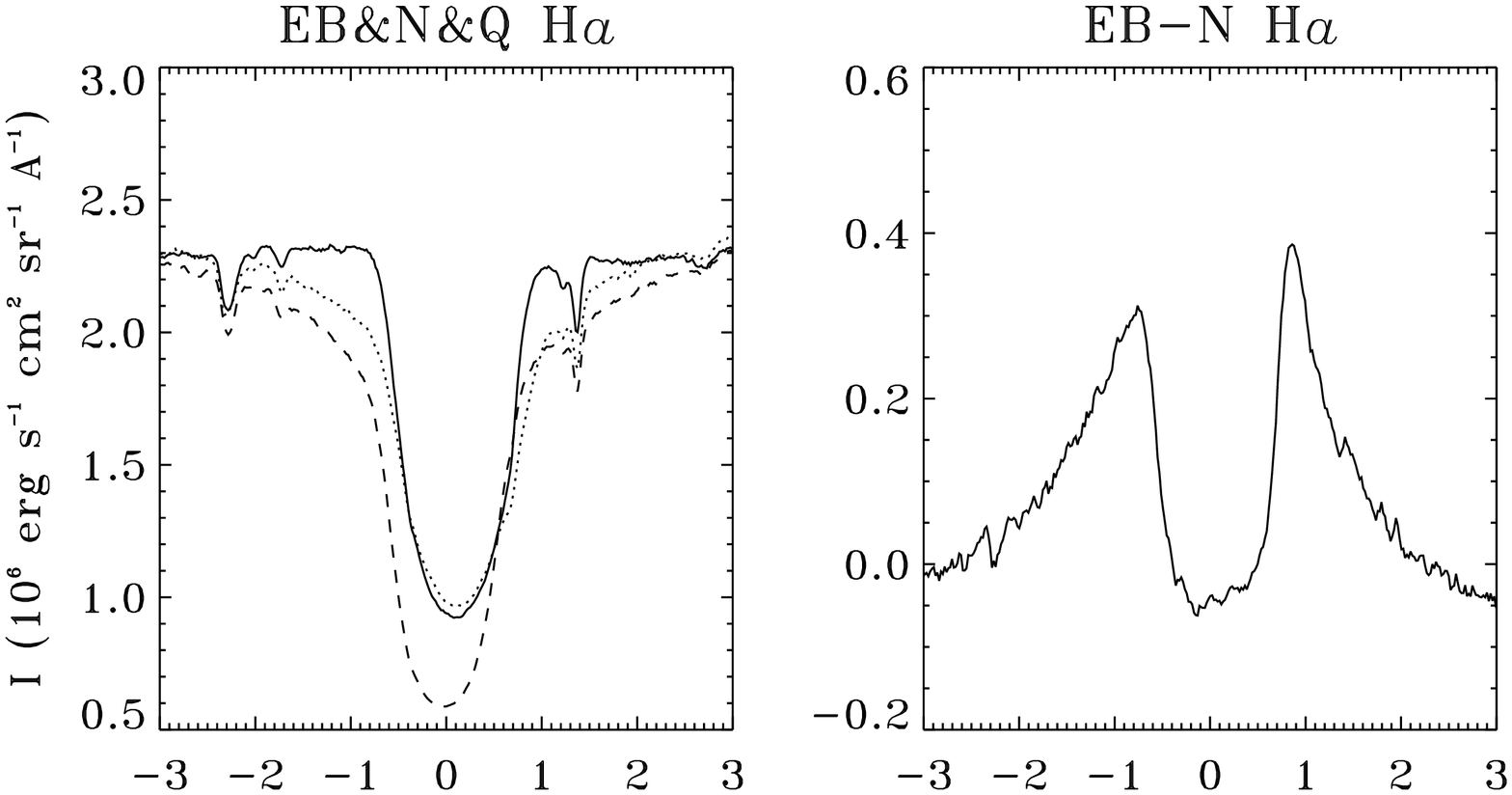}\\
\includegraphics[width=7.0cm,height=14.0cm,angle=90.0]{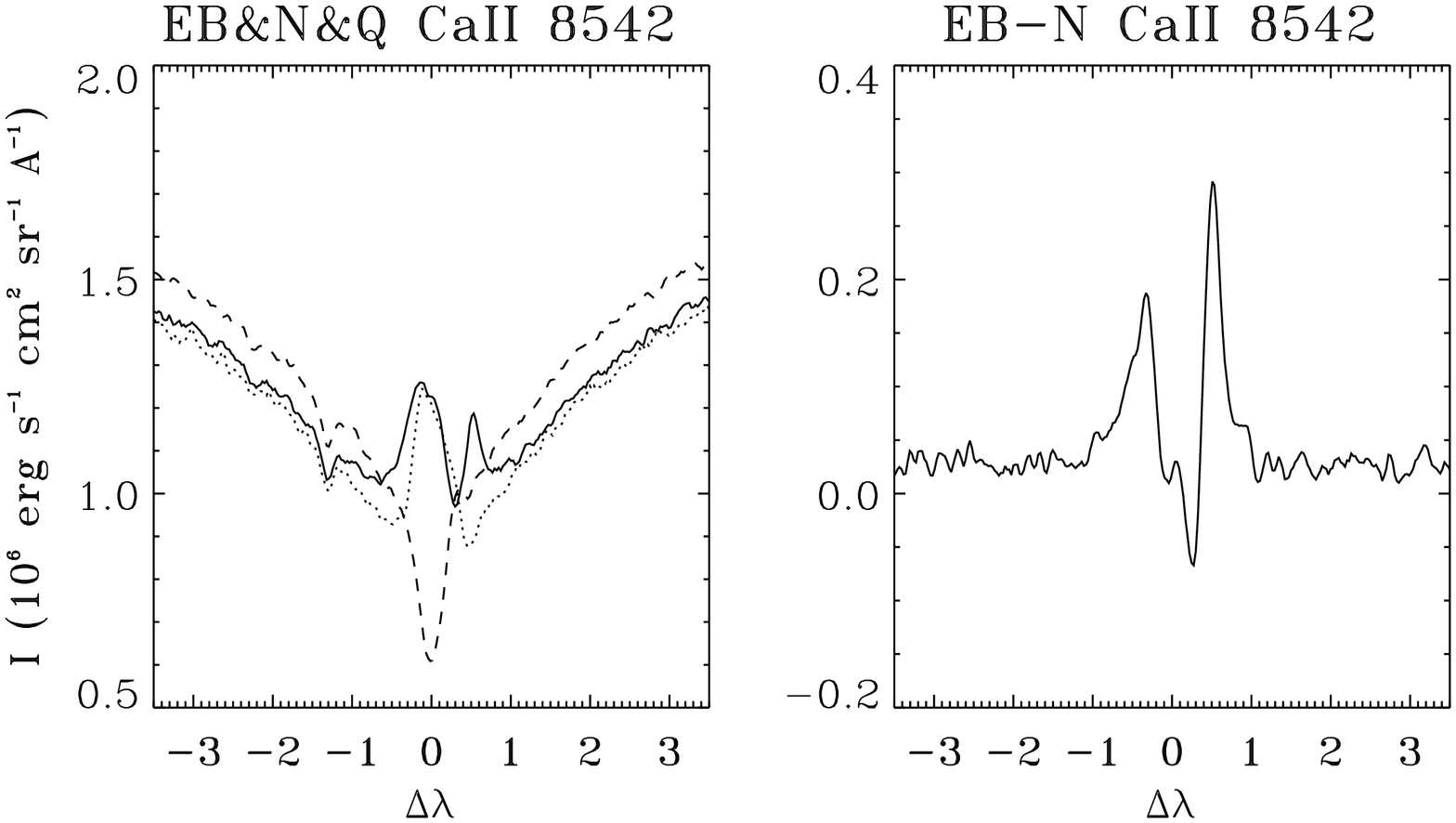}
\end{center}
 \caption{The left column shows the observed H$\alpha$ and \ion{Ca}{II} 8542 {\AA}
 line profiles of the EB No. 2 ({\it solid lines}), the line profiles of the nearby
 background (N; {\it dotted lines}), and those of the quiet-Sun (Q; {\it dashed lines}).
 The excess intensities $\Delta I$, i.e., the EB intensity with the nearby
 background being subtracted (EB-N), are shown in the two panels on the right
 column.}\label{fig1}
\end{figure}

By carefully checking the high-resolution spectra,
we found three well-observed small EBs during the observations. The criteria
for detecting EBs is the existence of excess emissions at the far
wings of the H$\alpha$ lines. Table \ref{tab1} lists some characteristics
of the three EBs, numbered No.1--No.3, including the time
when the EB intensity attains the maximum, $\Delta I$, duration $D$,
size, accompanied downward velocity, and $\Delta T$. Here
$\Delta I$ is the intensity difference between the EB H$\alpha$
peak (at $\sim -1$ {\AA}) and the nearby background, with a unit
of 10$^6$ erg s$^{-1}$ cm$^{-2}$ sr$^{-1}$ \AA$^{-1}$.
Note that the nearby background is a pre-heated area, not the quiet-Sun
further away, which has a lower intensity at the center of H$\alpha$ lines
as shown in Figure \ref{fig1}. The durations ($D$) of the EBs are
estimated by the scanning time during which the EB emission in
the far wing of the H$\alpha$ line can still be identified. The
sizes of EBs were determined by the full-width at half-maximum (FWHM)
of the lightcurves along the successive scan ($x$) and along
the slit ($y$) directions, when the
EBs can still be identified. A 2D Doppler velocity distribution
was recovered from spatially resolved H$\alpha$ line profiles
with the centroid method. $\Delta T$ is the peak
temperature difference between that in our EB semi-empirical
models (see \S\ref{sect4}) and that of the quiet-Sun model.

% ===============================================================================================
\begin{table*}
\begin{center}
\caption{Characteristics of the EBs}\label{tab1}
\end{center}
\begin{center}
\begin{tabular}{c c c c c c c}
\hline
 No. & Time & $\Delta I$ & Duration & Size ($x \times y$) & Downward $v_{\parallel}$ & $\Delta T$\\
     & (UT) & (10$^6$ erg s$^{-1}$ cm$^{-2}$ sr$^{-1}$ \AA$^{-1}$ )& (s) &(arcsec) & (km s$^{-1}$) & (K) \\
\hline
 1 &  17:03:04 &  $\geq$0.30 & 300   & 0.306$\times$ 0.310  &  4 & 2740\\
 2 &  17:09:55 &  $\geq$0.35 & 220   & 0.464$\times$ 0.368  &  5 & 2810\\
 3 &  17:22:03 &  $\geq$0.30 & 223   & 0.557$\times$ 0.765  &  5 & 2940\\

\hline
\end{tabular}

\end{center}
\end{table*}
%=================================================================================

Table \ref{tab1} shows that the sizes of the EBs are in the range of
0.3\arcsec\--0.8\arcsec\ with elongated structures, though
the No.1 EB has a more or less round structure. Moreover, all the EBs are
accompanied by downward flows.

\subsection{H$\alpha$ and \ion{Ca}{II} 8542 {\AA} line profiles of the EBs}

As an example, the H$\alpha$ and \ion{Ca}{II} 8542 {\AA} line profiles of the EB
No. 2 are shown as solid lines in the top-left panel and the bottom-left panel of
Figure  \ref{fig1}, respectively. For comparison, the dotted lines are the counterparts
of the nearby (N) background and the dashed lines indicate the line profiles
of the quiet-Sun regions (Q). The line profiles of the EB after subtracting those of the
nearby background, EB-N, are also shown in the top-right and bottom-right
panels. It can be seen that the EB-N profiles exhibit obvious excess emissions
at the blue and red wings, which is a typical characteristic of EBs. It
implies that the heating is significant in the solar lower atmosphere.
All these will be clearly seen in our computed semi-empirical atmospheric
models shown later in \S\ref{sect4}. It should be emphasized that comparing to the
quiet-Sun line profiles, there is an intensity enhancement at the EB H$\alpha$
line center. It implies that a heating still exists in the corresponding upper
 chromosphere. Note that the intensities at the blue and red wings are asymmetric, and
sometimes the blue wing is even stronger than the red wing.
The asymmetry might be produced by the dynamical processes in the EBs.

\subsection{2D magnetic and velocity maps around the EBs}
%Figure
\begin{figure}
\centerline{\includegraphics[width=11cm]{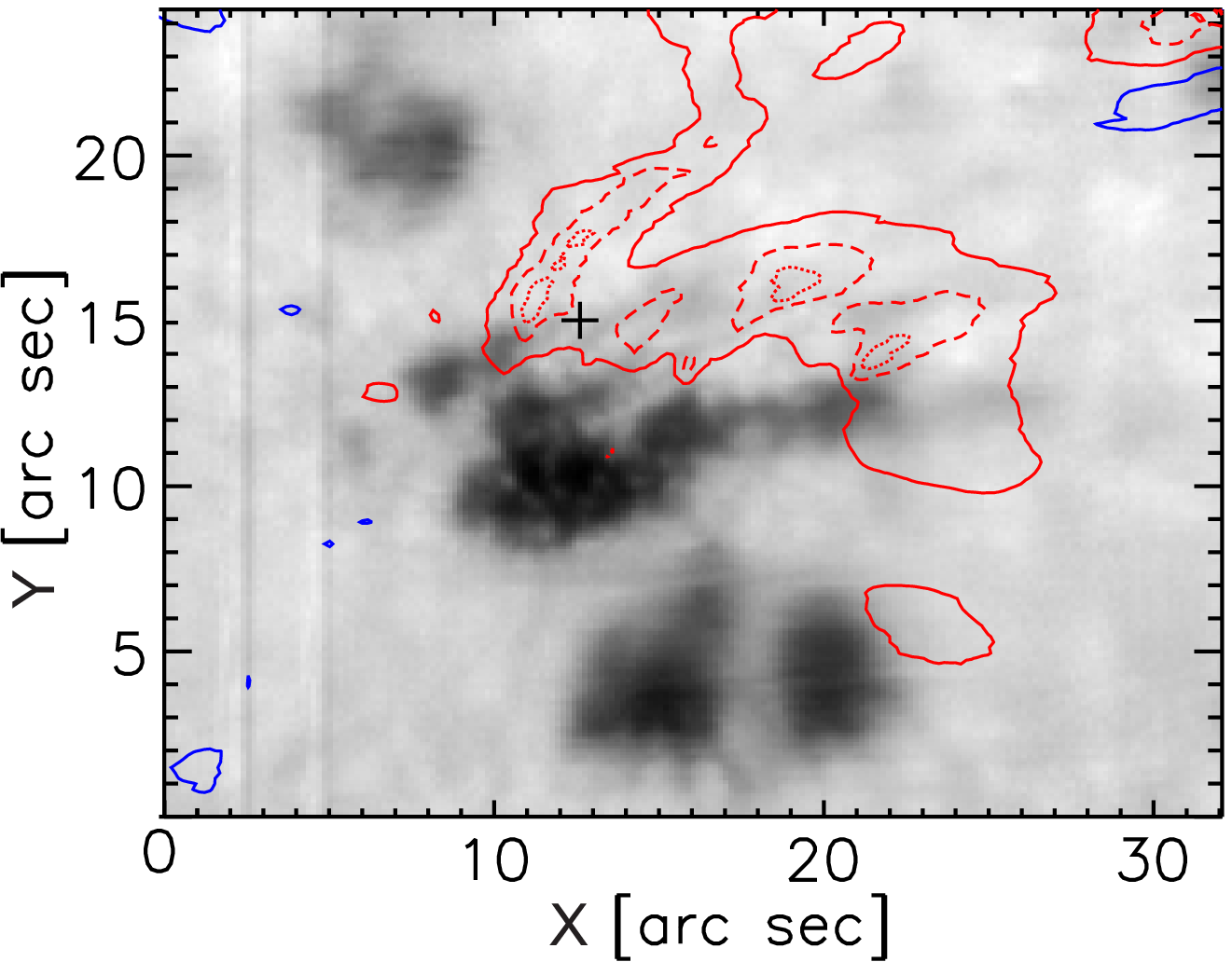}}
 \caption{Velocity distribution ({\it contours}) around 17:09:55 UT superposed
    on the H$\alpha$  blue wing ($-0.99$ \AA~ from the line center) image
    ({\it gray scale}). The cross sign pinpoints the EB No. 2. The contour levels
    of the Doppler velocity are -2, 3 (solid lines), 8 (dashed lines) and
    12 (dotted lines) km s$^{-1}$. The blue and red lines
    correspond to upward and downward velocities, respectively. Several
    vertical lines visible are the remains after the instrumentation
    artifact elimination.}\label{fig2}
\end{figure}

Figure \ref{fig2} displays the location of the EB No. 2 on the 2D FISS H$\alpha$
raster images taken at the far wing. The cross sign pinpoints the EB. The contours
show the velocity distribution around 17:09:55 UT with
the Doppler velocity levels of -2, 3, 8, 12 km s$^{-1}$. Figure \ref{fig2} shows
that the EB has a co-spatial relationship with
the downward mass motion measured in the H$\alpha$ profile.

\begin{figure}
\centerline{\includegraphics[width=11cm]{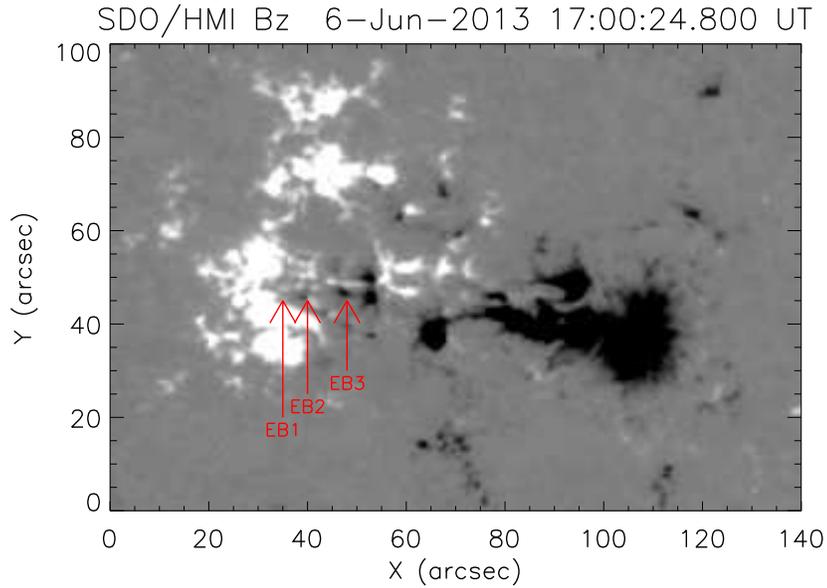}}
 \caption{Vertical component of the projection-corrected magnetic field.
    The arrows indicate the positions of the three EBs.}\label{fig3}
\end{figure}

\begin{figure}
\centerline{\includegraphics[width=13cm]{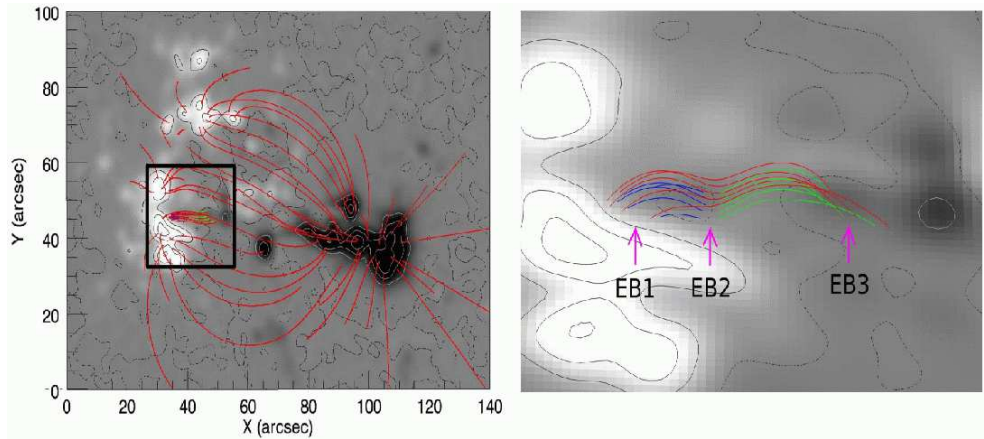}}
 \caption{Nonlinear force-free field extrapolated around the EBs. The
 black box in the left panel indicates the field of view of the right
 panel. Since the perspective of the right panel is different from that
 of the left one, the figure on the bottom is stretched. The left
 panel shows a large field of view, where the extrapolation is performed.
 The right panel focuses on a small field of view around the EBs.}\label{fig4}
\end{figure}

To derive the topology of magnetic field of the EBs, we make a nonlinear
force-free field (NLFF) extrapolation with an optimization method
(Wheatland et al.~\cite{Wheat00}; Wiegelmann et al.~\cite{ Wiegel04}).
The vector magnetogram is observed by the Helioseismic and Magnetic
Imager (\emph{HMI}; Scherrer et al.~\cite{Scher12}; Schou et al~\cite{Schou12})
on board the Solar Dynamics Observatory (SDO). First, we remove
the 180$^\circ$ ambiguity of the transverse components of the vector
magnetogram using the minimum energy method (Metcalf et al.~\cite{Met06};
Leka et al.~\cite{ Leka09}). Then, we correct the projection effect using
the method mentioned in Gary \& Hagyard(~\cite{Gary90}).
The vertical component of the projection corrected magnetic field is shown in
Figure \ref{fig3}. Next, a preprocessing technique (Wiegelmann et al.
~\cite{Wiegel06}) is applied to the vector magnetic field in the field of
view as shown in Figure \ref{fig3} to remove the net magnetic force and
torque. Finally, we derive the NLFF that is shown in Figure \ref{fig4}.
It shows that EB No.2 is co-spatial with a bipole, and EBs No.1 and No.3
are connected to the bipole with magnetic field lines as shown in the right
panel of Figure 4. We can see that the three EBs are located in the
region with parasitic magnetic elements showing mixed polarities, and are
connected with a series of magnetic field lines associated with bald patches,
where the magnetic field lines are tangential to the photosphere and concave up.

%Figure

%Figure

\subsection{Intensity evolution of the EBs}

\begin{figure}
\centerline{\includegraphics[width=11.0cm,height=8.0cm,angle=0.0]{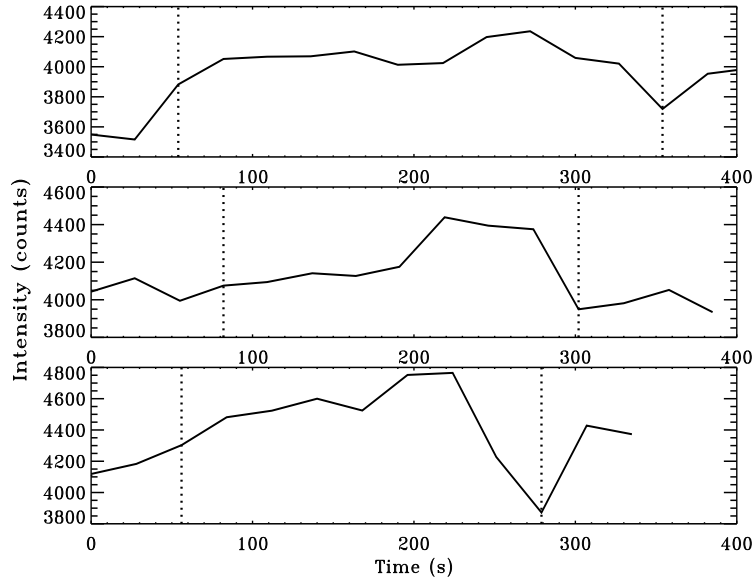}}
 \caption{Lightcurves of the three EBs  (in counts) at the
 far wings ($\sim$ -1 {\AA}) of their H$\alpha$ line profiles. From top to
 bottom:  the start time is 16:58:59 UT, 17:00:21 UT and 17:16:55 UT for
 EBs No. 1, No. 2, and No. 3, respectively. Two dashed lines mark the moments
 when the excess emission at the H$\alpha$ far wing appears and disappears,
respectively. The intensities plotted in this Figure are for one pixel where
the EBs have the peak intensities.
}\label{fig5}
\end{figure}

High spatial-resolution and high-cadence observations allow us to obtain
the intensity evolution at the H$\alpha$ far-wing ($\sim -1$~{\AA}) of
the three EBs as shown in Figure \ref{fig5}. Two dashed lines indicate the time
when the excess emission in the wing of H$\alpha$ lines appears and
disappears, respectively. The durations of the EBs listed in Table \ref{tab2}
are just taken as the time intervals between the appearance and the
disappearance of the excess emission of the EBs. According to the
property of the lightcurves, which are made from the single pixel
where the EB brightening attends the maximum, we can distinguish three
phases of the EB evolution: a pre-heating phase, when the intensity increased
slowly but continuously, and the excess emission at the far wing of the
H$\alpha$ lines is absent or very weak; a flaring phase, in which the
intensity increased quickly and the excess emission at the H$\alpha$
line wings attained the maximum at the peak time; and then a cooling
phase when the intensity decayed rapidly. The estimation of the magnetic
reconnection rate (see \S\ref{sect4}) indicates that the flaring phase
is produced by fast magnetic reconnection. It seems that the pre-heating
phase corresponds to a slow magnetic reconnection process, which produces
micro-turbulence. When it reaches a certain critical state, fast
magnetic reconnection commences. The cooling phase is short. Potentially, this is due
to the strong radiative loss in the solar lower atmosphere, where
the EBs occur.

%Figure

\section{Semi-empirical modelling of the small EBs}\label{sect4}

\subsection{Non-LTE computation of the semi-empirical models}

The semi-empirical atmospheric models of EBs can be computed by using
the H$\alpha$ and \ion{Ca}{II} 8542 {\AA} line profiles. We follow the
non-local thermal equilibrium (non-LTE) calculation method as described in the
paper of Fang et al.(~\cite{Fang06}).
We solve the statistical equilibrium equation, the transfer equation,
the hydrostatic equilibrium, and the particle conservation equations
iteratively. The relative difference of the mean intensity between
the last two iterations is less than 10$^{-7}$ and 10$^{-8}$ for
hydrogen and calcium atoms, respectively.

\begin{figure}
\centerline{\includegraphics[width=7.0cm,height=12.0cm,angle=90.0]{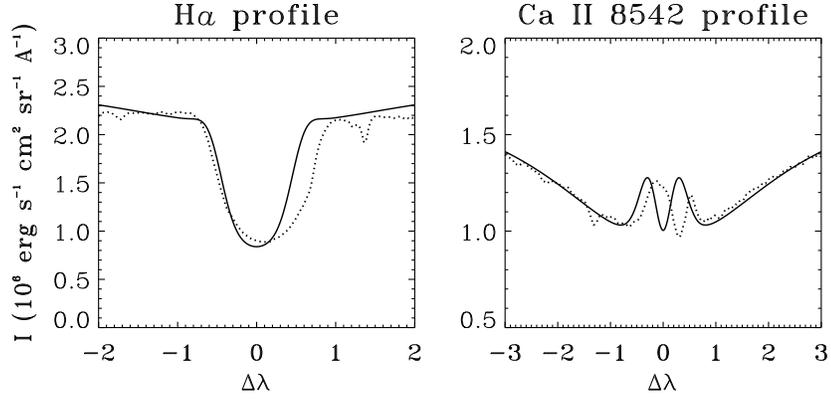}}
 \caption{Comparison between the observed ({\it dotted lines}) and non-LTE
 computed ({\it solid lines}) H$\alpha$ and \ion{Ca}{II} 8542 {\AA} line profiles
 for the EB No. 2. We can see the widening of the H$\alpha$ lines
 and the red shifting of both observed lines. These properties are
 discussed in the text.}\label{fig6}
\end{figure}

%Figure
\begin{figure}
\centerline{\includegraphics[height=11.0cm,angle=90.0]{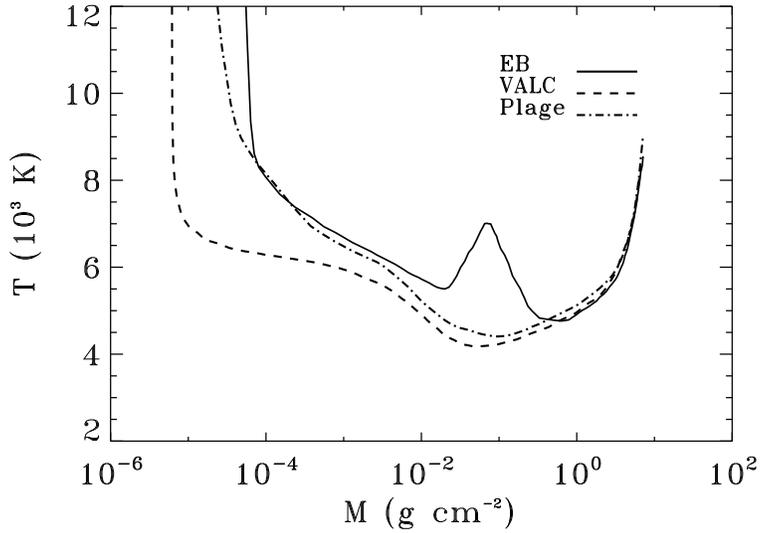}}
 \caption{Temperature distributions in the semi-empirical model
     of the EB No. 2 ({\it solid line}), compared to that of the plage
     model ({\it dashed-dotted line}) given by Fang et al.(~\cite{Fang01}), and
     that of the quiet-Sun model (i.e., the VALC model,  {\it dashed line}) given by
     Vernazza et al.(~\cite{Ver81}).}\label{fig7}
\end{figure}

As an example, Figure \ref{fig6} gives both the observed and computed H$\alpha$
and \ion{Ca}{II} 8542 {\AA} line profiles for the EB No. 2. It can be seen that
the modeled profiles can well match the observed ones, except two aspects:
(1) the computed H$\alpha$ line profile is narrower than the observed one.
This is probably due to the existence of turbulence in the EB heating
region, which might be caused during the magnetic reconnection and we
did not include in our computation; (2) The observed line profiles show
red shifts compared to the computed ones. This clearly indicates
that there is a downward motion also contributing
to the H$\alpha$ profile at this spatial position. We did not take
this effect into account in our modeling.
Figure \ref{fig7} gives the semi-empirical atmospheric model of the EB No. 2.
For comparison, we also plot the temperature distributions in the semi-empirical
model for plages (denoted by ``Plage'') as derived by Fang et al.(~\cite{Fang01})
and for the quiet-Sun model (denoted by ``VALC'') in Vernazza et al.(~\cite{Ver81}).
It can be seen there is a temperature increase, albeit weak, in the upper
chromosphere at the site of the EB, which is necessary to produce the intensity
increase of the EB at the center of the chromospheric lines compared to the
quiet-Sun one (see Figure \ref{fig1} in \S\ref{sect3}). The most distinct feature in
the semi-empirical model is an obvious heating around the temperature
minimum region and in the upper photosphere, which is responsible for
the excess emission at the far wings of the EB spectra. The maximum
temperature enhancements for the three EBs are in the range of
2700--3000 K (see Table \ref{tab1}), which is consistent with the
high-resolution observation by Berlicki et al.(~\cite{Berl10}), but much
higher than other previous values (e.g., Fang et al.~\cite{Fang06}).
The difference maybe comes from the fact that the previous observations
with a lower spatial resolution blended the EB intensity and the quiet-Sun one,
and resulted in weaker intensity than that in the high-resolution
observations. It is noted, however, that by use of the BBSO data,
Hong et al.(~\cite{Hong14}) found a lower temperature in EBs. This is because
they used an averaged (across an area over 2\arcsec) intensity line
profile.

%===================================================================

\subsection{Energy estimation of the EBs}

We use the method given in Fang et al.(~\cite{Fang06}) to estimate the
energy of EBs. It is hypothesized that the main heating regions of EBs are
in the lower chromosphere and the upper photosphere, so we can use
the following equation to estimate the radiative energy $E_r$ of EBs:

\begin{equation}
E_r = {D \over 2} A_{EB} \int_{h_1}^{h_2} {R_r}\, d {h} \,\,,
\end{equation}

\noindent
where the heating duration is assumed to be half of the EB
lifetime $D$. $A_{EB}$ is the area of the EB, which was
determined by the EB size ($x \times y$) listed in Table \ref{tab1}.
In Eq. (1), $h_1$ and $h_2$ are the lower and the upper
heights of the heated region, including the heating in
the chromosphere, $R_r$ the Non-LTE radiative losses in unit
of erg cm$^{-3}$ s$^{-1}$. Gan et al.(~\cite{Gan90}) provided a
semi-empirical formula for estimating $R_r$, but here we use
an improved empirical formula given by Jiang et al.(~\cite{Jiang10})
and shown as follows. This formula is more suitable for the
small-scale activities.

\begin{equation}
R_r= n_H n_e ({\alpha_1 (h)}+{\alpha_2 (h)}) f(T)\,,
\end{equation}

\noindent
where

$log ~{\alpha_1 (h)} = 1.745 \times 10^{-3} h
-4.739$\,,

$\alpha_2 (h) = 8.0 \times 10^{-2} e^{-3.701\times 10^{-2} h}$\,,

$f(T) = 4.533 \times 10^{-23} {(T/10^4)}^{2.874}$\,, \\

\noindent
where $h$ is the height in kilometers. To estimate the net
radiative energy $\Delta E$ of the EBs, we have to subtract
the radiative energy of the quiet-Sun ($E_Q$) from that of EBs ($E_r$):

\begin{equation}
\Delta E = E_r - E_Q \,\, .
\end{equation}

\noindent
$E_Q$ can be estimated by ${D \over 2} A_{EB} \int {R_Q} d{h}$,
where $\int {R_Q} dh$ is the radiative losses in the quiet-Sun
atmosphere. According to the result of Vernazza et al.(~\cite{Ver81}),
we take $\int {R_Q} d$h = 4.6$\times10^6$ erg cm$^{-2}$ s$^{-1}$.

The lower limit of the kinetic energy can be estimated by use of
the line-of-sight velocity near the EBs as

\begin{equation}
E_v = {1 \over 2} \times 1.4 m_H v_{\parallel}^2 f A_{EB}
\int_{h_3}^{h_4} n_H d {h} \,\,,
\end{equation}

\noindent
where $n_H$ is the hydrogen density, $f$ the fraction of the
mass involved in the motion. We assume $f= 0.1$ as in our
previous paper (Fang et al. ~\cite{Fang06}).  The coefficient 1.4 is
used for including the contribution from helium. $h_3$ and $h_4$
denote the lower and the upper heights of the EB main heating
region (corresponding to the temperature bump region, see
Figure \ref{fig7}), respectively. Actually, we take $h_3$ = $h_1$,
which are obtained from our semi-empirical models of the EBs.
Considering the rapid decrease of the hydrogen density with
height, we neglect the contribution from the higher layers.

Using Eqs. (3) and (4), the energies of the EBs can be estimated,
which are listed in Table \ref{tab2}. It can be seen that the total
energy is about 5$\times$10$^{25}$--3.0$\times$10$^{26}$ ergs,
which is in the lower limit range given by previous authors
(e.g. Georgoulis et al.~\cite{Geor02}; Fang et al.~\cite{Fang06}).
Considering the fact that these are three small EBs, it is reasonable.
In our cases, the radiative
and kinetic energies of the EBs are comparable.

\begin{table*}
\begin{center}
\caption{Energies and reconnection rates for the three EBs}\label{tab2}
\end{center}
\begin{center}
\begin{tabular}{c c c c c c c c c}
\hline
 EB & h1, h3 & h2 & h4 & $\Delta E$ & $E_v$ & $B$ & $R$ \\
    & (km) & (km) &(km)&   (erg)    & (erg) & (G) &   \\
\hline
No. 1&363&1926&760&3.12$\times 10^{25}$ &1.57$\times 10^{25}$ & 200--300 &0.12--0.036 \\
No. 2&337&2044&847&6.11$\times 10^{25}$ &7.14$\times 10^{25}$ & 200--300 &0.13--0.040 \\
No. 3&395&2386&804&2.02$\times 10^{26}$ &7.05$\times 10^{25}$ & 250--350 &0.19--0.071 \\
\hline
\end{tabular}
\end{center}
\end{table*}
%============================================================================

\subsection{Estimation of the magnetic reconnection rate}

Assuming that the thermal and kinetic energies of EBs come from the
dissipation of magnetic field during magnetic reconnection, we can
estimate the magnetic reconnection rate. Suppose that the heating
region of the EBs is the magnetic energy dissipating region, the
magnetic energy coming into the region per second is

\begin{equation}
E_{in} = {{V_{in} d l B^2} \over {\mu}} \,\, ,
\end{equation}

\noindent
where $d$ is the heated atmospheric height of the EBs and $l$ is the
 averaged apparent size of the EBs. We take $d =h4-h3$ and $l=(x + y)/2$.
$V_{in}$ is the inflow velocity, $B$ the magnetic field taken from
the \emph{HMI} observation, $\mu$ the
magnetic permeability.  If we take $E_{in}D = \Delta E + E_v$, we have

\begin{equation}
V_{in} = {{\mu (\Delta E + E_v)} \over {d l D B^2}}\,\, .
\end{equation}

 The Alfv\'en velocity $V_A = B / {\sqrt{\mu \rho}}$ and the density $\rho$
 can be obtained from our semi-empirical models. So we can obtain the
averaged reconnection rate, $R = V_{in}/V_A$, as follows:

\begin{equation}
R = {{\mu^{3/2} \rho^{1/2} (\Delta E + E_v)} \over {l d D B^3}}\,.
\end{equation}

The magnetic field strength can be obtained from the photospheric
magnetograms. The estimates of the reconnection rate $R$ for the
three EBs are listed in Table \ref{tab2}. It can be seen that $R$ is in
the range of 0.04--0.19, varying in different EBs and depending
on the magnetic field strength, but well in the regime of fast
magnetic reconnection (e.g., Priest \& Forbes~\cite{Priest00}). It implies
that the magnetic reconnection responsible for the EBs is Petschek-like
fast reconnection (Petschek~\cite{pets64}).

\section{Discussions and Conclusions}\label{sect5}

By using the FISS data of the 1.6 meter BBSO/NST telescope, we obtained
 high-resolution H$\alpha$ and \ion{Ca}{II} 8542 {\AA} spectra
of three well-observed small EBs. The high spatial resolution
data allow us to study the individual EBs without much mixing with
the surrounding region. The high temporal resolution data make
it possible to study the evolution of the EBs in detail, so as
to well clarify the different phases of the EBs.

It is shown that all the EBs are located near the parasitic
areas in the longitudinal magnetograms and are co-spatial to
mass motions of several km s$^{-1}$. Our NLFF extrapolation
clearly shows that the EBs appear at the bald patches and the
separatrices of the magnetic field, which
confirms the schematic model of Pariat et al.(~\cite{Par04}).
Checking the lightcurves of the EBs, the evolution of the EBs can
be divided into three phases: the pre-heating, flaring, and cooling
phases. The estimation of the magnetic reconnection rate of the EBs
indicates the occurrence of fast reconnection during the flaring
phase of the EBs. These facts imply that the EBs are caused by
Petscheck-type magnetic reconnection (e.g., H\'enoux et al.~\cite{Hen98};
Ding et al.~\cite{Ding98}), with
a rate similar to solar flares albeit with much smaller sizes. However,
compared to solar flares, the cooling phase of the EBs is much shorter.
It can be understood since in the solar lower atmosphere where
EBs occur, the radiative losses are much stronger than that in the
solar corona. Another reason is that part of the EB energy goes to
heat the upper chromosphere, so the cooling of the EBs should be quicker.

Using the Non-LTE theory, we computed the thermal semi-empirical models
for the three small EBs. Our results indicate that the required extra
temperature enhancement in the lower atmosphere is 2700--3000 K when
compared with the quiet-Sun model, as
shown in Figure \ref{fig7}. It can account for the excess emission at
the far wings of the chromospheric lines, which is the main spectral
feature of EBs. The temperature enhancement in our models is
larger than previous values given by some authors with lower
resolution observations. Such a result is not surprising. In fact,
with high spatial resolution observations, temperature increases
more than 2000 K have been reported (e.g., Georgoulis et al.
~\cite{Geor02}; Berlicki et al.~\cite{Berl10}). It is
probably that the temperature enhancement in many of the previous
works, i.e., $\sim$1000 K, was underestimated, or some of the events
are not real EBs (Rutten et al.~\cite{rutt13}). Another interesting
thing is that compared to the plage atmospheric model, there is also
a temperature enhancement in the EB upper chromosphere. It can be caused
by  jets or some kind of waves which are produced during the
magnetic reconnection process. Actually, in our previous numerical
simulations (Jiang et al.~\cite{Jiang10}; Xu et al.~\cite{Xu11}), the
temperature enhancement does appear in the upper chromosphere.

The semi-empirical models and the measured line-of-sight
velocities near the EBs are used to estimate both the radiative
and kinetic energies. Our results indicate that the total energy
of these three small EBs is about $5\times 10^{25}$--3.0$\times
10^{26}$ ergs .

Based on the analysis of the three small EBs, we draw the conclusions
as follows:

1. The thermal semi-empirical atmospheric models for the three small
EBs clearly show the heating bump around the temperature minimum
region. The temperature enhancement is about 2700--3000 K,
much higher than the values obtained previously with
lower-resolution spectral data.

2. All EBs are located near the parasitic magnetic areas in the
longitudinal magnetogram, and are accompanied by mass motions. Our
NLFF extrapolation shows that the EBs appear at the bald
patches and the separatrices of the magnetic field, which are
strongly suggestive of magnetic reconnection accounting for the
heating of EBs.

3. Combining the study of EB lightcurves and the
estimation of the magnetic reconnection rate, we propose a
three phase scenario for EBs brightenings: a pre-heating phase
which is probably produced by slow magnetic reconnection;
a flaring phase which is caused by fast reconnection,
and a following cooling phase. The excess emission at the
chromospheric line wings evidently appears in the flaring phase.

4. The radiative and kinetic energies are estimated. The
results indicate that the total energy of the EBs is about
5$\times$10$^{25}$--3.0$\times$10$^{26}$ ergs even for
these three small EBs with only sub-arcsecond sizes.

\begin{acknowledgements}
We would like to give our sincere gratitude to the staff
at the Big Bear Observatory of the New Jersey Institute
of Technology (NJIT) for their enthusiastic help during
CF's stay there. We also thanks a lot to the anonymous
Refree for his/her valuable comments and suggestions.
This work is supported by the National
Natural Science Foundation of China (NSFC) under the
grants 10878002, 10933003, 11025314, 10673004 and 11203014,
11103075 as well as NKBRSF under grants 2011CB811402 and
2014CB744203.  W. C. acknowledges the support of the US
NSF (AGS-0847126 and AGS-1250818) and NASA (NNX13AG14G).
\end{acknowledgements}

\end{document}